\RequirePackage{lineno}
\documentclass[aps,prd,superscriptaddress,
               preprintnumbers,amsmath,amssymb,nofootinbib,12pt]{revtex4}

\usepackage{ulem}
\usepackage{graphicx}
\usepackage{wasysym}
\usepackage{dcolumn}
\usepackage{color}
\usepackage{fancyvrb}
\usepackage{gensymb}
\usepackage{caption}
\usepackage{subcaption}
\graphicspath{{ps}}

\usepackage{setspace}

\begin{document}


\vspace*{-3\baselineskip}
\resizebox{!}{3cm}{\includegraphics{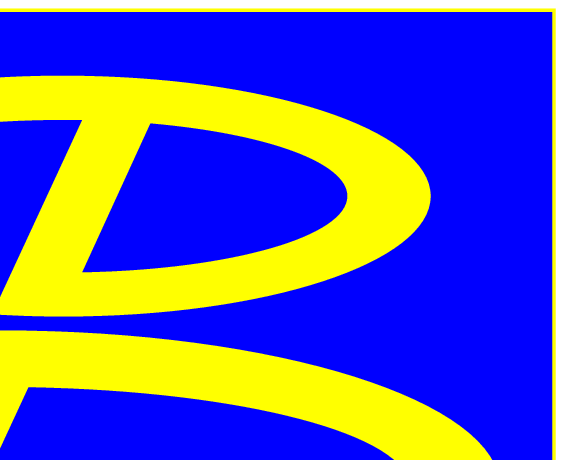}}
\vskip -3cm
\noindent
\hspace*{4.5in}BELLE Preprint 2014-8 \\
\hspace*{4.5in}KEK Preprint 2014-2 \\

\vskip 2cm
\begin{center}
{\Large \bf Search for $B^+ \to e^+ \nu$ and $B^+ \to \mu^+ \nu$ decays using hadronic tagging}\\
\end{center}
\singlespacing
\begin{center}
{\large Abstract}
\end{center}

{\indent \small We present a search for the rare leptonic decays 
$B^+ \to e^+ \nu_e$ and $B^+ \to \mu^+ \nu_\mu$, using the full 
$\Upsilon(4S)$ data sample of $772\times 10^6$ $B\bar{B}$ pairs 
collected with the Belle detector at the $\textsc{KEKB}$ 
asymmetric-energy $e^+ e^-$ collider. One of the $B$ mesons 
from the $\Upsilon(4S) \to B\bar{B}$ decay is fully
reconstructed in a hadronic mode, while the recoiling
side is analyzed for the signal decay. We find no evidence of a signal
in any of the decay modes. Upper limits of the corresponding branching
fractions are determined as ${\cal B}(B^+ \to e^+ \nu_e)<3.5\times                                     
10^{-6}$ and ${\cal B}(B^+ \to \mu^+ \nu_\mu)<2.7 \times 10^{-6}$ at
90$\%$ confidence level.\\
}


\noaffiliation
\affiliation{University of the Basque Country UPV/EHU, 48080 Bilbao}
\affiliation{Beihang University, Beijing 100191}
\affiliation{University of Bonn, 53115 Bonn}
\affiliation{Budker Institute of Nuclear Physics SB RAS and Novosibirsk State University, Novosibirsk 630090}
\affiliation{Faculty of Mathematics and Physics, Charles University, 121 16 Prague}
\affiliation{University of Cincinnati, Cincinnati, Ohio 45221}
\affiliation{Deutsches Elektronen--Synchrotron, 22607 Hamburg}
\affiliation{Justus-Liebig-Universit\"at Gie\ss{}en, 35392 Gie\ss{}en}
\affiliation{Gifu University, Gifu 501-1193}
\affiliation{Hanyang University, Seoul 133-791}
\affiliation{University of Hawaii, Honolulu, Hawaii 96822}
\affiliation{High Energy Accelerator Research Organization (KEK), Tsukuba 305-0801}
\affiliation{Hiroshima Institute of Technology, Hiroshima 731-5193}
\affiliation{IKERBASQUE, Basque Foundation for Science, 48011 Bilbao}
\affiliation{Indian Institute of Technology Bhubaneswar, Satya Nagar 751007}
\affiliation{Indian Institute of Technology Guwahati, Assam 781039}
\affiliation{Indian Institute of Technology Madras, Chennai 600036}
\affiliation{Institute of High Energy Physics, Chinese Academy of Sciences, Beijing 100049}
\affiliation{Institute of High Energy Physics, Vienna 1050}
\affiliation{Institute for High Energy Physics, Protvino 142281}
\affiliation{INFN - Sezione di Torino, 10125 Torino}
\affiliation{Institute for Theoretical and Experimental Physics, Moscow 117218}
\affiliation{J. Stefan Institute, 1000 Ljubljana}
\affiliation{Kanagawa University, Yokohama 221-8686}
\affiliation{Institut f\"ur Experimentelle Kernphysik, Karlsruher Institut f\"ur Technologie, 76131 Karlsruhe}
\affiliation{Kavli Institute for the Physics and Mathematics of the Universe (WPI), University of Tokyo, Kashiwa 277-8583}
\affiliation{Department of Physics, Faculty of Sciences, King Abdulaziz University, Jeddah 21589}
\affiliation{Korea Institute of Science and Technology Information, Daejeon 305-806}
\affiliation{Korea University, Seoul 136-713}
\affiliation{Kyungpook National University, Daegu 702-701}
\affiliation{\'Ecole Polytechnique F\'ed\'erale de Lausanne (EPFL), Lausanne 1015}
\affiliation{Faculty of Mathematics and Physics, University of Ljubljana, 1000 Ljubljana}
\affiliation{Luther College, Decorah, Iowa 52101}
\affiliation{University of Maribor, 2000 Maribor}
\affiliation{Max-Planck-Institut f\"ur Physik, 80805 M\"unchen}
\affiliation{School of Physics, University of Melbourne, Victoria 3010}
\affiliation{Moscow Physical Engineering Institute, Moscow 115409}
\affiliation{Moscow Institute of Physics and Technology, Moscow Region 141700}
\affiliation{Graduate School of Science, Nagoya University, Nagoya 464-8602}
\affiliation{Kobayashi-Maskawa Institute, Nagoya University, Nagoya 464-8602}
\affiliation{Nara Women's University, Nara 630-8506}
\affiliation{National Central University, Chung-li 32054}
\affiliation{National United University, Miao Li 36003}
\affiliation{Department of Physics, National Taiwan University, Taipei 10617}
\affiliation{H. Niewodniczanski Institute of Nuclear Physics, Krakow 31-342}
\affiliation{Nippon Dental University, Niigata 951-8580}
\affiliation{Niigata University, Niigata 950-2181}
\affiliation{Osaka City University, Osaka 558-8585}
\affiliation{Pacific Northwest National Laboratory, Richland, Washington 99352}
\affiliation{Panjab University, Chandigarh 160014}
\affiliation{Peking University, Beijing 100871}
\affiliation{University of Pittsburgh, Pittsburgh, Pennsylvania 15260}
\affiliation{University of Science and Technology of China, Hefei 230026}
\affiliation{Seoul National University, Seoul 151-742}
\affiliation{Soongsil University, Seoul 156-743}
\affiliation{Sungkyunkwan University, Suwon 440-746}
\affiliation{School of Physics, University of Sydney, NSW 2006}
\affiliation{Department of Physics, Faculty of Science, University of Tabuk, Tabuk 71451}
\affiliation{Tata Institute of Fundamental Research, Mumbai 400005}
\affiliation{Excellence Cluster Universe, Technische Universit\"at M\"unchen, 85748 Garching}
\affiliation{Toho University, Funabashi 274-8510}
\affiliation{Tohoku Gakuin University, Tagajo 985-8537}
\affiliation{Tohoku University, Sendai 980-8578}
\affiliation{Tokyo Institute of Technology, Tokyo 152-8550}
\affiliation{Tokyo Metropolitan University, Tokyo 192-0397}
\affiliation{Tokyo University of Agriculture and Technology, Tokyo 184-8588}
\affiliation{CNP, Virginia Polytechnic Institute and State University, Blacksburg, Virginia 24061}
\affiliation{Wayne State University, Detroit, Michigan 48202}
\affiliation{Yamagata University, Yamagata 990-8560}
\affiliation{Yonsei University, Seoul 120-749}

  \author{Y.~Yook}\affiliation{Yonsei University, Seoul 120-749} 
  \author{Y.-J.~Kwon}\affiliation{Yonsei University, Seoul 120-749} 
  \author{A.~Abdesselam}\affiliation{Department of Physics, Faculty of Science, University of Tabuk, Tabuk 71451} 
  \author{I.~Adachi}\affiliation{High Energy Accelerator Research Organization (KEK), Tsukuba 305-0801} 
  \author{S.~Al~Said}\affiliation{Department of Physics, Faculty of Science, University of Tabuk, Tabuk 71451}\affiliation{Department of Physics, Faculty of Sciences, King Abdulaziz University, Jeddah 21589} 
  \author{K.~Arinstein}\affiliation{Budker Institute of Nuclear Physics SB RAS and Novosibirsk State University, Novosibirsk 630090} 
  \author{D.~M.~Asner}\affiliation{Pacific Northwest National Laboratory, Richland, Washington 99352} 
  \author{V.~Aulchenko}\affiliation{Budker Institute of Nuclear Physics SB RAS and Novosibirsk State University, Novosibirsk 630090} 
  \author{T.~Aushev}\affiliation{Institute for Theoretical and Experimental Physics, Moscow 117218} 
  \author{R.~Ayad}\affiliation{Department of Physics, Faculty of Science, University of Tabuk, Tabuk 71451} 
% ...
\collaboration{The Belle Collaboration}

\maketitle
\tighten
{\renewcommand{\thefootnote}{\fnsymbol{footnote}}}
\setcounter{footnote}{0}

\newpage

The purely leptonic decay $B^+ \to \ell^+ \nu_\ell$, where $\ell$ 
represents $e$, $\mu$ or $\tau$~\footnote{Charge-conjugate modes 
are implied throughout this paper unless stated otherwise.}, proceeds 
via annihilation of the $B^+$~meson's constituent quarks into a 
positively charged lepton and a neutrino of the 
same generation. In the Standard Model (SM), this annihilation is 
mediated by a $W^+$ boson. The branching fraction is
calculated~\cite{SM} by
\begin{equation} {\cal B}(B^+ \to \ell^+ \nu_\ell) =
  {{G_{F}^{2}m_{B}m_{\ell}^{2}}\over{8\pi}}\left(
    1-{{m_{\ell}^{2}}\over{m_{B}^{2}}} \right)^2
  f_{B}^{2}|V_{ub}|^{2}\tau_{B}, \end{equation} 
where $G_{F}$ is the
Fermi coupling constant, $m_{\ell}$ is the mass of the charged lepton,
$m_{B}$ is the mass of the $B^+$~meson, $\tau_{B}$ is the
$B^{+}$~meson lifetime, $V_{ub}$ is an element of the
Cabibbo-Kobayashi-Maskawa~(CKM) matrix~\cite{CKM} governing the weak
transition from the $b$ to the $u$ quark and $f_{B}$ is the $B$ decay
constant. The estimated branching fractions using
$|V_{ub}|=(3.51_{-0.14}^{+0.15})\times 10^{-3}$~\cite{PDG} from a fit
to the full CKM unitarity triangle and $f_B = 186\pm 4 ~{\rm
  MeV}$~\cite{FBQCD} from lattice QCD calculations and the world
average for all other parameters~\cite{PDG} are ${\cal B}(B^+ \to e^+
\nu_e)=(7.9_{-0.7}^{+0.8})\times 10^{-12}$, ${\cal B}(B^+ \to \mu^+
\nu_\mu)=(3.4\pm 0.3)\times 10^{-7}$, and ${\cal B}(B^+ \to \tau^+
\nu_\tau)=(7.5\pm 0.7)\times 10^{-5}$.

The $B^+ \to \tau^+ \nu_\tau$ mode has been measured previously by the
Belle~\cite{taube} and $\textsc{Babar}$~\cite{tauba} experiments,
resulting in a combined branching fraction of $(1.05\pm 0.25)\times
10^{-4}$~\cite{PDG}. Due to the relatively small expected branching
fractions, owing to helicity suppression in the SM,
observation of the $B^+ \to e^+ \nu_e$ and $B^+ \to
\mu^+ \nu_\mu$ decay modes remains a
challenge. Currently, the most stringent upper limits
  of these decays are ${\cal B}(B^{+} \to e^{+}\nu_e) < 9.8 \times
  10^{-7}\textrm{~\cite{Belle2007}}$ and ${\cal B}(B^{+} \to
  \mu^{+}\nu_\mu) < 1.0 \times 10^{-6}\textrm{~\cite{Babar2009}}$ at
  90$\%$ confidence level~(C.L.).

The $B^+ \to \ell^+ \nu_\ell$ decays provide an
excellent probe for new physics (NP), thanks to the small theoretical
uncertainty in the SM branching fractions. For instance, in NP
scenarios containing hypothetical particles such as the charged Higgs
in 2-Higgs doublet models (type-II)~\cite{2HDM} or the minimal 
supersymmetric model (MSSM)~\cite{MSSM} or leptoquarks~\cite{LQ}, the
branching fractions of the $B^+ \to \ell^+ \nu_\ell$ decays can be
greatly enhanced.

Moreover, it has been suggested that the relative
branching fractions of $B^+ \to \ell^+ \nu_\ell$ to $B^+
  \to \ell'^+ \nu_{\ell'}$, $R^{\ell\ell'}= {\cal B}( B^+ \to \ell^+\nu_\ell) /
  {\cal B} (B^+\to \ell'^+\nu_{\ell'})$ where $\ell \neq \ell'$, 
can be used to test the minimal flavor violation (MFV) hypothesis. 
In NP models with MFV~\cite{MLFV}, the ratios $R^{\ell\ell'}$ are 
expected to be nearly unmodified from SM expectations. 
However, in the framework of a grand unified theory (GUT) model, 
$B^+ \to \ell^+\nu_{\ell'}$ decays can additionally 
contribute as to increase the ratios $R^{e\mu}$ and $R^{e\tau}$ by 
more than 1 order of magnitude above SM expectations~\cite{FI}. 
It has been also suggested that, in a general MSSM
model at large $\tan\beta$~\cite{tanbeta} with heavy
squarks~\cite{Lepuniv}, the ratios $R^{e\tau}$ and $R^{\mu\tau}$ can
deviate from SM expectations. Therefore measurements of $B^+ \to e^+
\nu_e$ and $B^+ \to \mu^+ \nu_\mu$ combined with the existing $B^+ \to
\tau^+ \nu_\tau$ determination can provide significant constraints on NP. 

In this paper, we present a search for the previously unobserved $B^+
\to \ell^+ \nu_\ell$ decays, using the hadronic tagging method, where
$\ell$ stands for $e$ or $\mu$~\footnote{From this point and on,
$\ell$ represents $e$ and $\mu$ only.}. In the hadronic tagging method, we
fully reconstruct one of the $B$ mesons from the $\Upsilon(4S) \to
B\bar{B}$ decay in a hadronic mode and then select the $B^+ \to \ell^+
\nu_\ell$ signal from the rest of the event. The existing upper limits
on the branching fraction determined using the hadronic tagging method
are ${\cal B}(B^{+} \to e^{+}\nu_e) < 5.2 \times 10^{-6}$ and ${\cal
  B}(B^{+} \to \mu^{+}\nu_\mu) < 5.6 \times
10^{-6}\textrm{~\cite{Babar2008}}$ at 90$\%$
C.L. These are not as stringent as the limits 
mentioned above which were obtained by the so-called untagged 
method.  But the hadronic tagging analysis is complementary to the 
untagged one in that it has a better sensitivity to discern new 
physics effects if it occurs. 

By not explicitly reconstructing a $B$ meson, the untagged method does
not fully utilize the information from the accompanying $B$~meson
decay. While it leads to higher signal selection efficiencies, it
suffers from a substantially higher background level. This could lead
to ambiguities with other processes having similar decay signatures in
case a signal is observed far in excess of the SM expectation. For
instance, if an unknown heavy neutrino $\nu_h$~\cite{Nuh} appears in
the $B^+\to e^+\nu_h$ decay, it will be nearly impossible to
distinguish it from the known process, $B^+ \to e^+ \nu_e$, because of
the limited kinematic precision of the untagged method.

In the hadronic tagging method, by fully reconstructing one $B$ meson 
($B_{\rm tag}$), we have the best possible knowledge on the kinematics 
of the signal $B$ meson ($B_{\rm sig}$) in the event. 
This enables a precise measurement of the missing four-momentum of 
the neutrino in the $B^+ \to \ell^+ \nu_\ell$ decays. 
As a result, the momentum of the charged lepton in
the $B^+ \to \ell^+ \nu_\ell$ signal can be determined with an
order-of-magnitude higher resolution compared to the untagged
method~\cite{Belle2007}. This results in a very strong background
suppression and provides an extra constraint for identifying the
nature of the undetected particle.

The data sample used in this analysis was collected with the Belle
detector~\cite{Belle} at the KEKB asymmetric-energy $e^+ e^-$
collider~\cite{KEKB}. The sample corresponds to an integrated
luminosity of 711~$\rm fb^{\rm -1}$ or $772\times10^6$ $B\bar{B}$
pairs, collected on the $\Upsilon(4S)$ resonance at a center-of-mass
(CM) energy ($\sqrt{s}$) of $10.58~{\rm GeV}$. 

The Belle detector is a large-solid-angle spectrometer that consists
of a silicon vertex detector (SVD), a 50-layer central drift chamber
(CDC), aerogel threshold Cherenkov counters (ACC), an array of a
barrel-like arrangement of time-of-flight scintillation counters
(TOF), and an electromagnetic calorimeter comprised of 8736 CsI(Tl)
crystals (ECL) located inside a superconducting solenoid coil that
provides a 1.5~T magnetic field. An iron flux return located outside
of the coil is instrumented to detect $K_L^0$ mesons and to identify
muons (KLM). 

Electron identification is based on the ratio between the cluster
energy in the ECL and the track momentum from the CDC $(E/p)$, the
specific ionization $dE/dx$ in the CDC, the position and shower shape
of the cluster in the ECL and the response from the ACC. Muon
identification is based on the hit position and the penetration depth
in the KLM. In the momentum range of interest in this analysis, the
electron (muon) identification efficiency is above $90\%$ and the
hadron fake rate is under $0.5\%~(5\%)$. A more detailed description
can be found elsewhere~\cite{LID}. 


The $B_{\rm tag}$ candidates are reconstructed in 615 exclusive
charged $B$~meson decay channels with a reconstruction algorithm based
on a hierarchical neural network~\cite{EKP}. To compensate for the
difference between the MC and data in the $B_{\rm tag}$ tagging 
efficiency ($\epsilon_{\textrm{tag}}$) due to uncertainties in
branching fractions and dynamics of hadronic modes, we apply a
correction obtained from a control sample
study~\cite{Ulnu} in which the signal-side $B$~meson
decays via five semileptonic $B^+$ decay modes:
  $B^+\to \bar{D}^0(K^+\pi^-)\ell^+\nu_\ell$, $B^+\to
  \bar{D}^0(K^+\pi^-\pi^0)\ell^+\nu_\ell$, $B^+\to
  \bar{D}^0(K^+\pi^-\pi^-\pi^+)\ell^+\nu_\ell$, $B^+\to
  \bar{D}^{*0}[\bar{D}^{0}(K^+\pi^-)\pi^0]\ell^+\nu_\ell$, and
  $\bar{D}^{*0}[\bar{D}^{0}(K^+\pi^-)\gamma]\ell^+\nu_\ell$.
The MC efficiency is corrected according to the $B_{\rm tag}$ decay
mode as well as the output of the hadronic tagging algorithm
($\it{o}_{\rm tag}$) on an event-by-event basis. The ${\it{o}_{\rm
    tag}}$ distribution peaks near zero for combinatorial or continuum
backgrounds, and near one for well reconstructed $B_{\rm tag}$
candidates.

\begin{figure*}[htb!]
	\begin{subfigure}[b]{0.48\textwidth}
		\includegraphics[width=\textwidth]{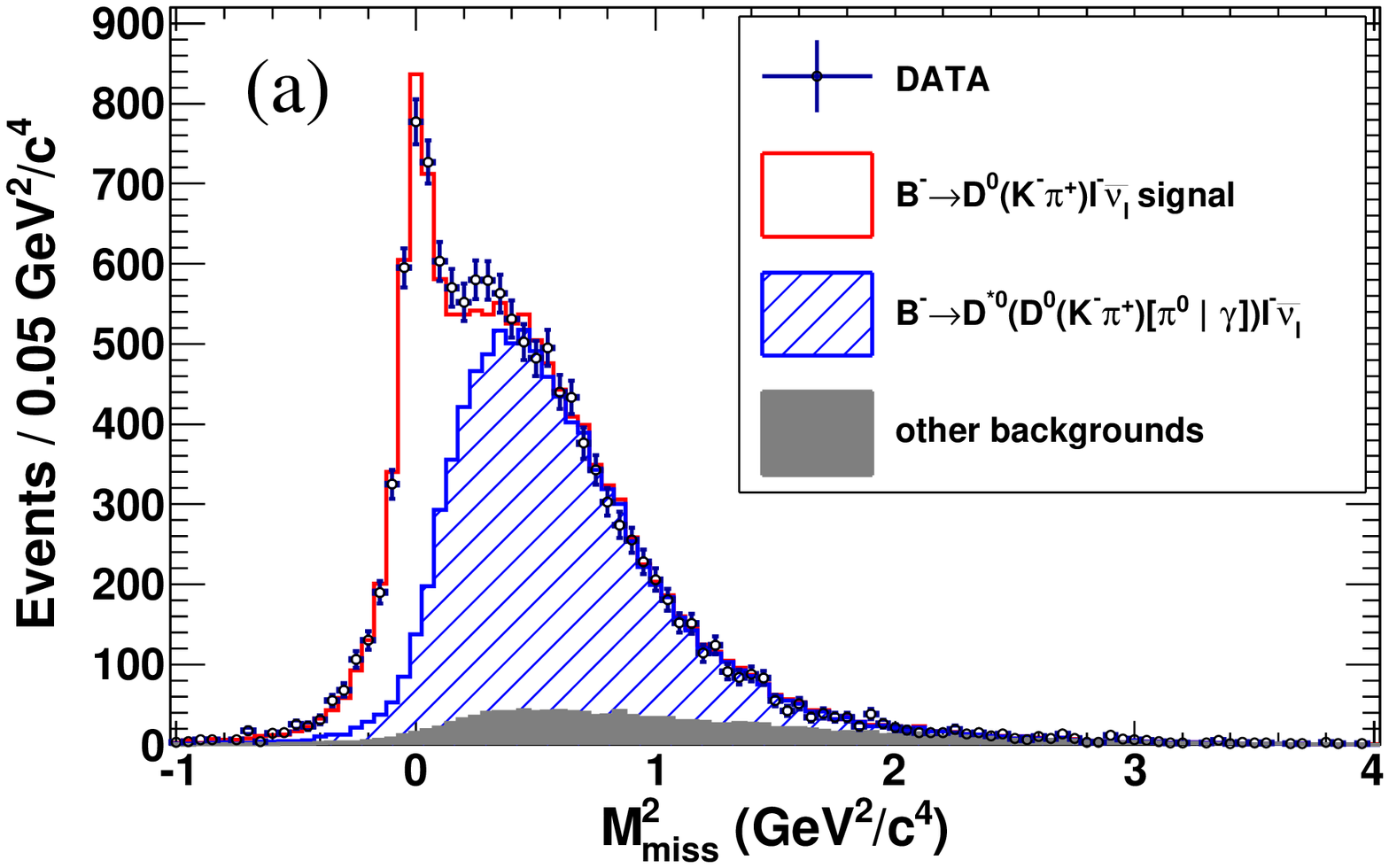}
		\label{d0kp}
	\end{subfigure}
	\begin{subfigure}[b]{0.48\textwidth}
		\includegraphics[width=\textwidth]{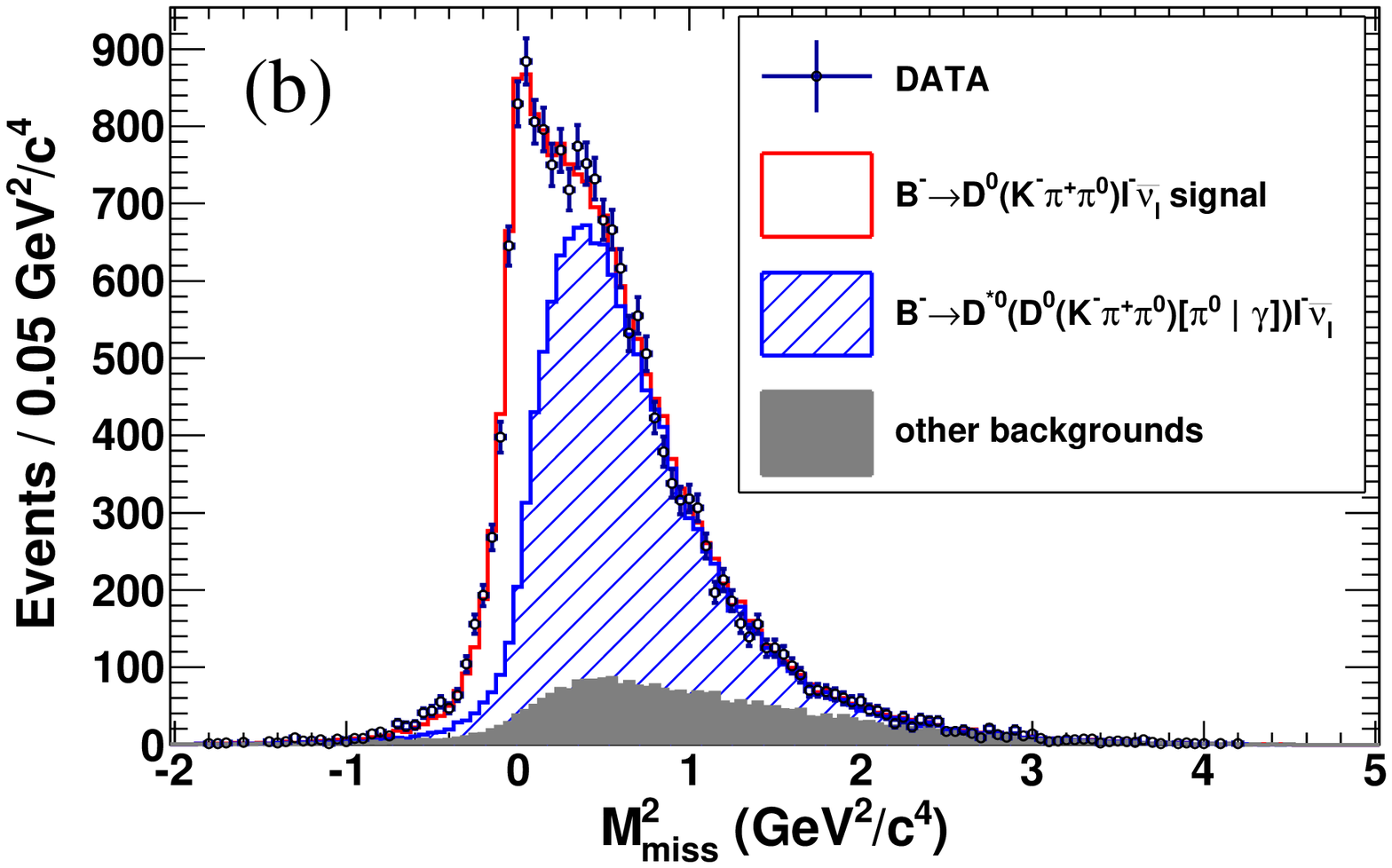}
		\label{d0kpp}
	\end{subfigure}
	\begin{subfigure}[b]{0.48\textwidth}
		\includegraphics[width=\textwidth]{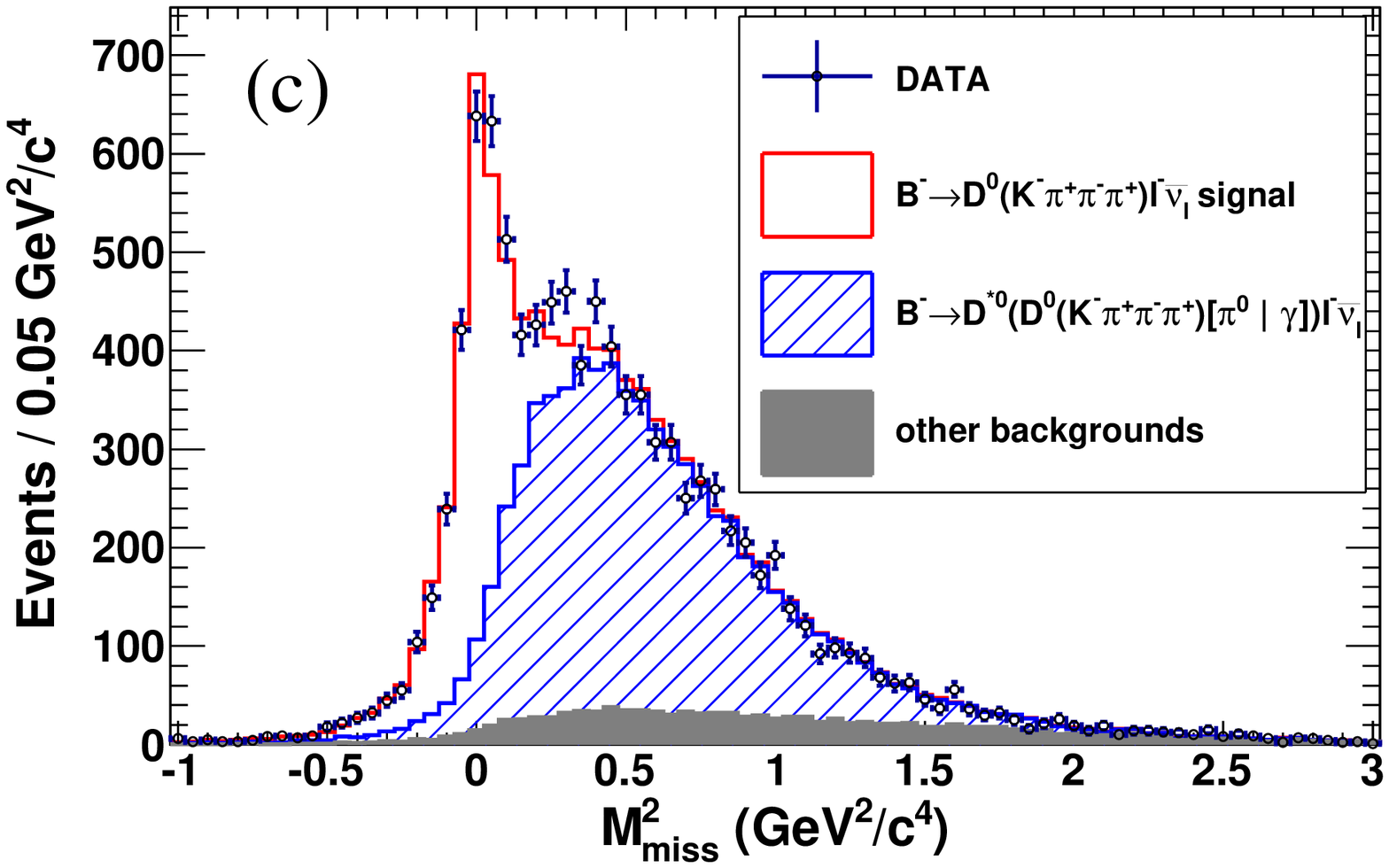}
		\label{d0kppp}
	\end{subfigure}
	\begin{subfigure}[b]{0.48\textwidth}
		\includegraphics[width=\textwidth]{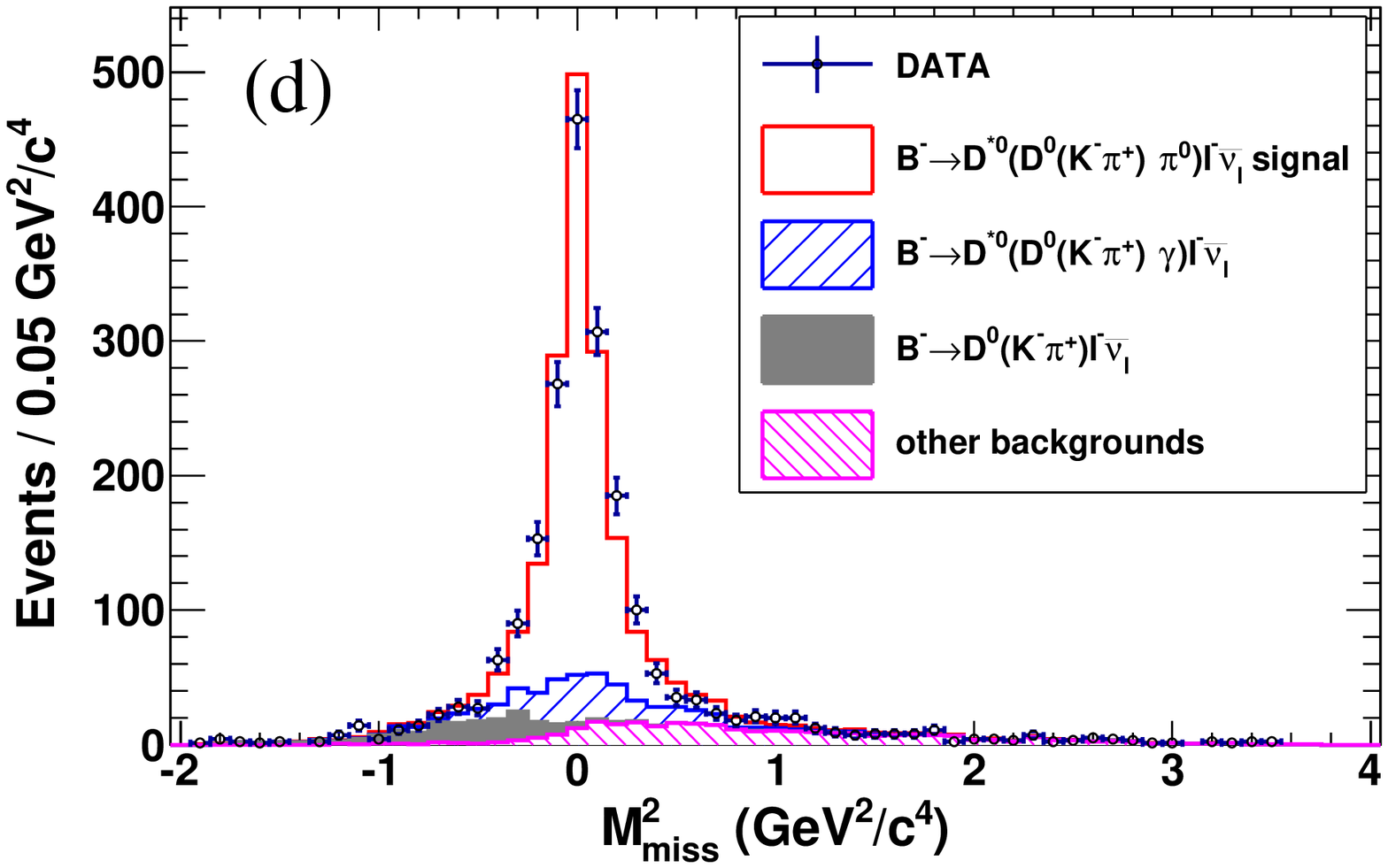}
		\label{dst0kpp}
	\end{subfigure}
	\begin{subfigure}[b]{0.48\textwidth}
		\includegraphics[width=\textwidth]{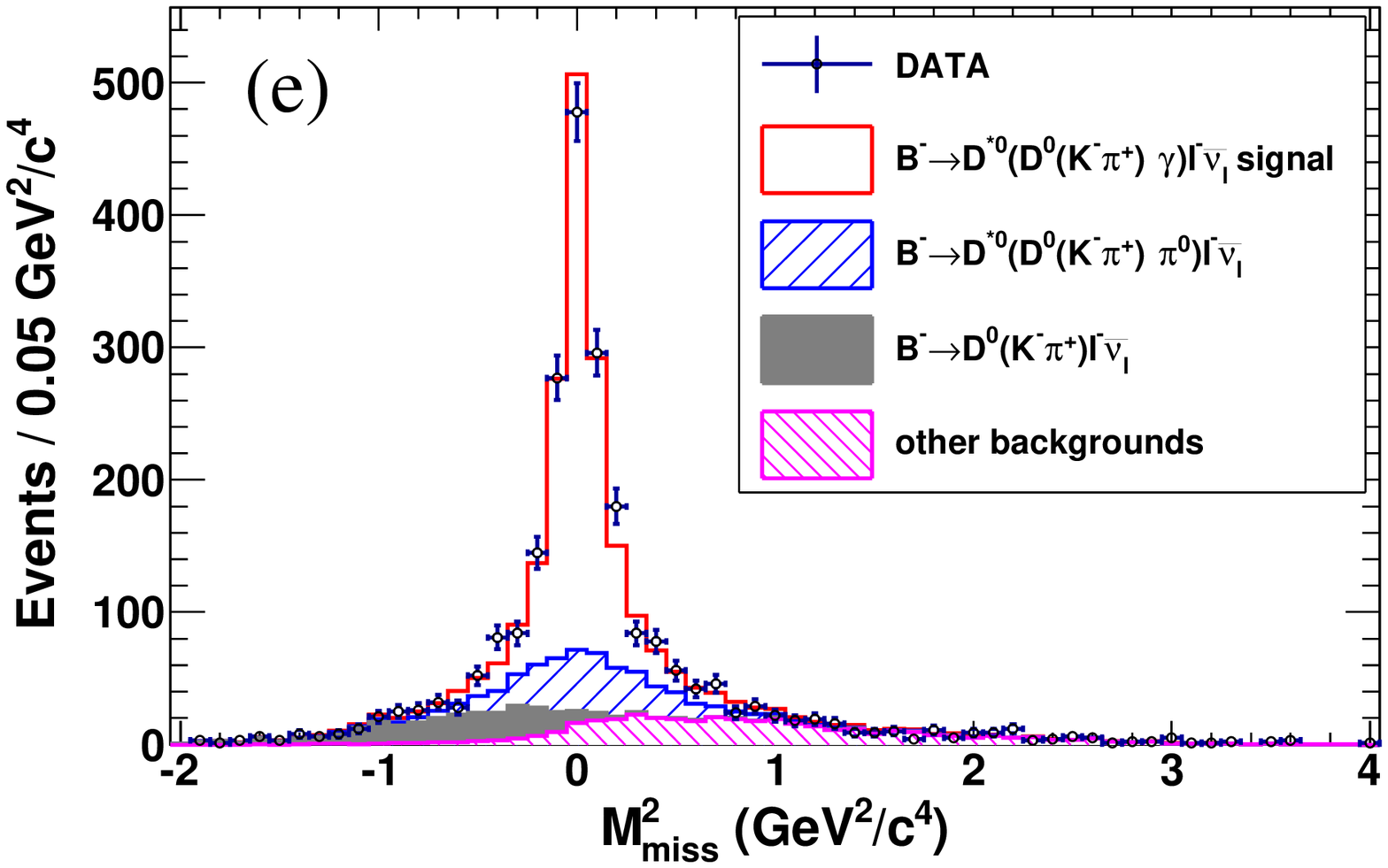}
		\label{dst0kpg}
	\end{subfigure}	
\caption{Fits to the $M_{\rm miss}^2$ distribution in data 
using the tagging efficiency corrected MC histogram templates in each of the 
(a) $B^+\to \bar{D}^0(K^+\pi^-)\ell^+\nu_\ell$,
(b) $B^+\to \bar{D}^0(K^+\pi^-\pi^0)\ell^+\nu_\ell$, 
(c) $B^+\to \bar{D}^0(K^+\pi^-\pi^-\pi^+)\ell^+\nu_\ell$, 
(d) $B^+\to \bar{D}^{*0}(\bar{D}^{0}[K^+\pi^-]\pi^0)\ell^+\nu_\ell$,
and (e) $B^+\to \bar{D}^{*0}(\bar{D}^{0}[K^+\pi^-]\gamma)\ell^+\nu_\ell$ control sample modes.
The other backgrounds component as listed in the legends is consisted of $b\to c$ decays, 
$e^+e^-\to q\bar{q}~(q = u,d,s,c)$ decays, and $b \to u\ell^-\bar{\nu}_\ell$ decays.}
\label{tagfit}
\end{figure*}

The correction factor for each $B_{\rm tag}$ decay mode is determined
by the comparison of the number of events in MC and data from a
one-dimensional binned maximum likelihood (ML) fit using histogram
templates~\cite{Barlow}, which take account of both the data and MC
statistical uncertainty, to the distribution of the square of the
missing particle's undetected four-momentum~($M_{\rm miss}^2$).  Here
$M_{\rm miss}^2$ is expected to peak near zero for correctly
reconstructed $B^+ \to \bar{D}^{(*)0} \ell^+ \nu_\ell$ events in which
the only missing particle is a massless neutrino as displayed in
Figure~\ref{tagfit}. The correction factor is then obtained from each
of the five control samples and we apply the averaged factor in our
analysis. The systematic uncertainty of the $\epsilon_{\rm tag}$
correction is estimated including the statistical precision of the
correction, the uncertainty of the branching fraction of the control
sample modes~\cite{PDG}, the effects of background-modeling to the 
$M_{\rm miss}^2$ fitting, and the uncertainty due to the 
particle identification used in reconstructing
the $D^{*0}$ mesons obtained by studying the $D^{*+}\to D^{0}\pi^+$
decay followed by the $D^{0}\to K^-\pi^+$ decay. Including the
systematic uncertainty, we finally obtain the correction factor as
$0.71\pm 0.05$ in both the $B^+
\to e^+ \nu_e$ and $B^+ \to \mu^+ \nu_\mu$ signal MC samples, with a
total fractional uncertainty of 6.4\% to
the correction factor.

In $42\%$ of the events for both the $B^+ \to e^+ \nu_e$ and $B^+ \to
\mu^+ \nu_\mu$ signal MC samples, we find multiple $B_{\rm
  tag}$ candidates. In such cases, we select the $B_{\rm tag}$
candidate with the highest $\it{o}_{\rm tag}$. To ensure a well
reconstructed $B_{\rm tag}$ candidate, we further require
${\it{o}_{\rm tag}}$, the energy difference, $\Delta E = E_{B_{\rm
    tag}}^{\, *} - {\sqrt{s}}/2$, and the beam-constrained-mass
$M_{\rm bc} = \sqrt{s/4 - |{\vec{p}_{B_{\rm tag}}^{\, *}}|^2}$, to
satisfy ${\it{o}_{\rm tag}} > 0.0025$, $|\Delta E|<0.05~{\rm GeV}$,
and $5.27~{\rm GeV}/c^2 < M_{\rm bc} <5.29~{\rm GeV}/c^2$, where
$E_{B_{\rm tag}}^{\, *}$ and $\vec{p}_{B_{\rm tag}}^{\, *}$ are the
$B_{\rm tag}$ energy and momentum, respectively, in the CM frame. The
efficiencies of this $B_{\rm tag}$ reconstruction procedure on events
containing signal decays are: $\epsilon_{\textrm{tag}} = 0.29$
$\pm\ 0.02$~\% for the $B^+ \to e^+ \nu_e$ and
$\epsilon_{\textrm{tag}} = 0.30$ $\pm\ 0.02$~\% for
the $B^+ \to \mu^+ \nu_\mu$. These $\epsilon_{\textrm{tag}}$ values
include the correction factor described above.

On the $B_{\rm sig}$ side, we require exactly one remaining track in
the detector and that it be identified as an electron or a
muon.  Since the signal mode is a two-body decay of a
  $B^+$ meson, the lepton momentum in the rest frame of the $B_{\rm
    sig}$ ($p_\ell^B$) peaks sharply around $2.64~{\rm GeV}/c$.  To
  utilize this feature while keeping a sideband available for
  background estimation, the lepton candidates are initially required
  to have a momentum above $1.8~{\rm GeV}/c$ in the laboratory frame.
They are also required to satisfy
$|dz|<1.5~{\rm cm}$ and $dr<0.05~{\rm cm}$, where $dz$ and $dr$ are
impact parameters of the track along the beam direction and in the
perpendicular plane, respectively.

To suppress the continuum background $(e^+ e^- \to q\bar{q}~[q =
u,d,s,c])$, we use the event shape difference between $B\bar{B}$
events and continuum. Since each $\Upsilon(4S)$ decays nearly at rest,
the decay products of the resulting $B\bar{B}$ pair have a spherical
event shape. On the other hand, continuum event shapes tend to be
two-jet-like. We define $\theta_{\rm T}$ as the angle between the
momentum of the signal lepton and the unit vector $\hat{n}$ that
maximizes $\Sigma_{i} {|\hat{n}\cdot\vec{p}_{i}|}/{|\vec{p}_{i}|}$,
where the index $i$ runs over all particles used for $B_{\rm tag}$
reconstruction. We require $\cos\theta_{\rm T}<0.9$ and
$\cos\theta_{\rm T}<0.8$ for $B^+ \to e^+ \nu_e$ and $B^+ \to \mu^+
\nu_\mu$, respectively. In the muon mode, we expect a larger continuum
background compared to the electron mode due to the higher hadron
misidentification rate. Therefore, we apply a more stringent
$\cos\theta_{\rm T}$ criterion for this mode.

For signal events, we expect no detectable particles left after
removing the signal lepton and the particles associated with the
$B_{\rm tag}$. Therefore, there should be no
extra energy deposits in the ECL except for the small contributions
from split-off showers and beam background. We define the extra energy
($E_{\rm ECL}$) as the sum of the energy from the neutral clusters not
associated with $B_{\rm tag}$ or the signal lepton deposited in the
ECL. In the $E_{\rm ECL}$ calculation, minimum thresholds of $50~{\rm
  MeV}$ for the barrel $(32.2^{\circ}<\theta<128.7^{\circ})$,
$100~{\rm MeV}$ for the forward end-cap
$(12.4^{\circ}<\theta<31.4^{\circ})$, and $150~{\rm MeV}$ for the
backward end-cap $(130.7^{\circ}<\theta<155.1^{\circ})$ of the
calorimeter are required, where $\theta$ is the cluster's polar angle
relative to the beam direction~\cite{Belle}. Higher thresholds are
applied for the end-cap regions due to the severity of beam background
there. We require $E_{\rm ECL}<0.5~{\rm GeV}$ for both $B^+ \to e^+
\nu_e$ and $B^+ \to \mu^+ \nu_\mu$.

We identify signal events with 
$\mathbf{\it p}_\ell^B$. By studying the signal MC samples, 
we demand that each signal event satisfies 
$2.6~{\rm GeV}/c < \mathbf{\it p}_\ell^B < 2.7~{\rm GeV}/c$
for both $B^+ \to e^+ \nu_e$ and $B^+ \to \mu^+ \nu_\mu$.

Dominant backgrounds arise from decays with neutral particles not
detected or used in the reconstruction of the $B_{\rm tag}$ and a high
momentum track that falls in the $\mathbf{\it p}_\ell^B$ signal
region. For the $B^+ \to e^+ \nu_e$,
$B^{+}\to\pi^{+}K^{0}$, $B^{+}\to\ell^{+}\nu_{\ell}\gamma$, and
$B^{+}\to\pi^{0}\ell^{+}\nu_{\ell}$ decays in our sample constitute
100$\%$ of the background events in the $\mathbf{\it p}_\ell^B$ signal
region. For the $B^+ \to \mu^+ \nu_\mu$,
$B^{+}\to\pi^{+}K^{0}$, $B^{+}\to K^{+}\pi^{0}$,
$B^{+}\to\ell^{+}\nu_{\ell}\gamma$, and
$B^{+}\to\pi^{0}\ell^{+}\nu_{\ell}$ decays constitute $84.7\%$ of the
background events in the $\mathbf{\it p}_\ell^B$ signal region with
the remainder coming from all other $b\to u\ell^{-}\bar{\nu}_{\ell}$
decays. For an accurate modeling of the background probability density
function (PDF) near the $\mathbf{\it p}_\ell^B$ signal region, we
generate dedicated MC samples for $B^{+}\to\pi^{+}K^{0}$, $B^{+}\to
K^{+}\pi^{0}$, $B^{+}\to\ell^{+}\nu_{\ell}\gamma$, and
$B^{+}\to\pi^{0}\ell^{+}\nu_{\ell}$ decays. For the
$B^{+}\to\ell^{+}\nu_{\ell}\gamma$ process, which has not been
observed yet, we assume a branching fraction of ${\cal B}(B^+\to
\ell^+\nu_\ell\gamma) = 5\times 10^{-6}$~\cite{Lnugam}.

We define the sideband of the $\mathbf{\it p}_\ell^B$ as $2.0~{\rm
  GeV}/c < \mathbf{\it p}_\ell^B < 2.5~{\rm GeV}/c$. The $\mathbf{\it
  p}_\ell^B$ sideband is dominated by the $b\to c$ and $b\to
u\ell^{-}\bar{\nu}_{\ell}$ decays. Out of all background events in the
$\mathbf{\it p}_\ell^B$ sideband, each $b\to c$ and $b\to
u\ell^{-}\bar{\nu}_{\ell}$ decay contributes $55\%~(60\%)$ and
$39\%~(34\%)$ for the $B^+ \to e^+ \nu_e~(B^+ \to \mu^+ \nu_\mu)$. 
The remaining $6\%$ of the background in the
$\mathbf{\it p}_\ell^B$ sideband originates from the
$B^{+}\to\ell^{+}\nu_{\ell}\gamma$ decay and the $b \to s,d$ processes
aside from $B^{+}\to\pi^{+}K^{0}$ or $B^{+}\to K^{+}\pi^{0}$ for both
searches. $B^{+}\to\bar{D}^{*0}\ell^{+}\nu_{\ell}$ and
$B^{+}\to\bar{D}^{0}\ell^{+}\nu_{\ell}$ decays are found to be
composing the $b\to c$ decays for the $B^+ \to e^+ \nu_e~(B^+ \to
\mu^+ \nu_\mu)$ at rates of $67\%~(64\%)$ and
$24\%~(21\%)$, respectively, and are treated separately from the other
$b\to c$ decays.

Continuum events are found to be negligible in both the $\mathbf{\it
  p}_\ell^B$ sideband and $\mathbf{\it p}_\ell^B$ signal regions. 


We calculate the branching fraction as
\begin{equation} {\cal B}(B^+ \to \ell^+ \nu) = {{N_{\rm obs} - N_{\rm
        exp}^{\rm bkg}}\over{2\cdot \epsilon_{\rm s} \cdot N_{B^{+}
        B^{-} }}},\end{equation} where $N_{\rm obs}$ is the observed
yield of the data sample in the $\mathbf{\it p}_\ell^B$ signal region,
$N_{\rm exp}^{\rm bkg}$ is the expected number of background events in the
$\mathbf{\it p}_\ell^B$ signal region, $\epsilon_{\rm s}$ is the total
signal selection efficiency, and $N_{B^{+} B^{-}}$ is the number of
$\Upsilon(4S) \to B^{+} B^{-}$ events in the data sample. Using ${\cal
  B}(\Upsilon(4S) \to B^{+} B^{-}) = 0.513\pm 0.006$~\cite{PDG}, we
estimate $N_{B^{+} B^{-}}$ as $(396\pm7)\times10^6$.

We obtain $N_{\rm exp}^{\rm bkg}$ by fitting the
  $\mathbf{\it p}_\ell^B$ sideband of the data sample, with a PDF
  obtained from the background MC. We then estimate the expected
  background yield in the $\mathbf{\it p}_\ell^B$ signal region from the
  ratio of the fitted background MC yields in the $\mathbf{\it
    p}_\ell^B$ sideband and the $\mathbf{\it p}_\ell^B$ signal
  region.

The systematic uncertainties on $N_{\rm exp}^{\rm bkg}$ are estimated
according to the uncertainties in the background PDF parameters, the
branching fraction of background decays, and the statistics of the
data sample in the $\mathbf{\it p}_\ell^B$ sideband. We vary each
source in turn by its uncertainty ($\pm 1\sigma$) and the resulting
deviations in $N_{\rm exp}^{\rm bkg}$ are added in
quadrature. To calculate the effect of the branching fraction uncertainties 
of the background modes, we refer to the experimental measurements~\cite{PDG} for the
$B^{+}\to\bar{D}^{(*)0}\ell^{+}\nu_{\ell}$,
$B^{+}\to\pi^{0}\ell^{+}\nu_{\ell}$, $B^{+}\to\pi^{+}K^{0}$, and
$B^{+}\to K^{+}\pi^{0}$ modes, and vary each branching 
fraction one by one from the world-average value by its error. For
the $B^{+}\to\ell^{+}\nu_{\ell}\gamma$, an uncertainty of $\pm 50\%$
is applied. For modes where a clear estimate of the background level
is not available, we assume a conservative branching fraction
uncertainty of $_{-50}^{+100}\%$. The values of $N_{\rm exp}^{\rm
  bkg}$ and their uncertainties for both $B^+ \to e^+ \nu_e$ and $B^+
\to \mu^+ \nu_\mu$ decays are listed in Table~\ref{res}.

\begin{table}[htb]
\begin{center}
  \caption{Summary of the signal selection efficiency
      ($\epsilon_{\rm s}$), the number of events observed in the
      $p_\ell^B$ signal region ($N_{\rm obs}$), and the expected
      background yield in the $p_\ell^B$ signal region ($N_{\rm
        exp}^{\rm bkg}$) for the $B^+ \to \ell^+ \nu_\ell$ search.}
\label{res}
\begin{tabular}
{@{\hspace{0cm}}l@{\hspace{0.25cm}}
  @{\hspace{0.25cm}}c@{\hspace{0.25cm}}
  @{\hspace{0.25cm}}c@{\hspace{0.25cm}}
  @{\hspace{0.25cm}}c@{\hspace{0.25cm}}} 
\hline\hline
{Mode} & {$\epsilon_{\rm s}~[\%]$} & {$N_{\rm obs}$} & {$N_{\rm exp}^{\rm bkg}$}\\
\hline
{$B^+ \to e^+ \nu_e$}  & $0.086\pm 0.007$ & {$0$} & {$0.10\pm 0.04$} \\
{$B^+ \to \mu^+ \nu_\mu$} & $0.102\pm 0.008$ & {$0$}
 & {$0.26_{-0.08}^{+0.09}$} \\ 
\hline\hline
\end{tabular}
\end{center}
\end{table}

\begin{table}[htb]
\begin{center}
\caption{Summary of multiplicative systematic uncertainties related to
  the $\epsilon_{\rm s}N_{B^+ B^-}$ calculation, in percent.} 
\label{sys}
\begin{tabular}
{@{\hspace{0cm}}l@{\hspace{0.5cm}}
  @{\hspace{0.5cm}}c@{\hspace{0.5cm}}
  @{\hspace{0.5cm}}c@{\hspace{0cm}}} 
\hline\hline
{Source}        & {$B^+ \to e^+ \nu_e$} & {$B^+ \to \mu^+ \nu_\mu$}\\
\hline
{$N_{B^+ B^-}$} & {$1.8$}             & {$1.8$}\\
{Lepton ID}     & {$2.0$}             & {$2.3$}\\
{MC statistics} & {$1.4$}             & {$1.3$}\\
{Tracking efficiency} & {$0.35$}      & {$0.35$}\\
{$\epsilon_{\textrm{tag}}$ correction} & {$6.4$}
  & {$6.4$}\\
{$\mathbf{\it p}_\ell^B$ Shape}        & {$3.6$}       & {$3.6$}\\
\hline
{Total}         & {$8.0$} & {$8.0$}\\
\hline\hline
\end{tabular}
\end{center}
\end{table}

The efficiencies $\epsilon_{\rm s}$ are $0.086\pm 0.007$ and 
$0.102\pm 0.008$ for $B^+ \to e^+ \nu_e$ and $B^+ \to
\mu^+ \nu_\mu$, respectively, as summarized in Table~\ref{res}.  The
uncertainties of $\epsilon_{\rm s}$ are calculated from the following
sources: lepton identification, signal MC statistical error, track
finding uncertainties of the signal lepton, $\epsilon_{\textrm{tag}}$
correction, and $\mathbf{\it p}_\ell^B$ shape.
 
The lepton identification efficiency correction is estimated by
comparing the efficiency difference between the data and MC using
$\gamma\gamma\to e^+e^-/\mu^+\mu^-$ processes, from which we obtain a
$2.0\%$ uncertainty for $B^+ \to e^+ \nu_e$ and $2.3\%$ for $B^+ \to
\mu^+ \nu_\mu$. The uncertainty due to signal MC statistics is $1.4\%$
for $B^+ \to e^+ \nu_e$ and $1.3\%$ for $B^+ \to \mu^+ \nu_\mu$. The
track-finding uncertainty is obtained by studying the partially
reconstructed $D^{*+}\to D^{0}\pi^+$, $D^{0}\to K_{S}^{0} \pi^+\pi^-$,
and $K_{S}\to \pi^+\pi^-$ decay chain, where one of the $K_{S}^{0}$
daughters is not explicitly reconstructed. We compare, between data
and MC, the efficiency of finding the $K_S^0$ daughter pion which is
not explicitly used in the partial $D^*$ reconstruction and estimate a
contribution of $0.35\%$ uncertainty for both $B^+ \to e^+\nu_e$ and
$B^+\to \mu^+\nu_\mu$ modes. We also include the 6.4\% 
$\epsilon_{\textrm{tag}}$ correction uncertainty mentioned earlier.

\begin{figure}[htb]
\begin{center}
\includegraphics[width=0.45\textwidth]{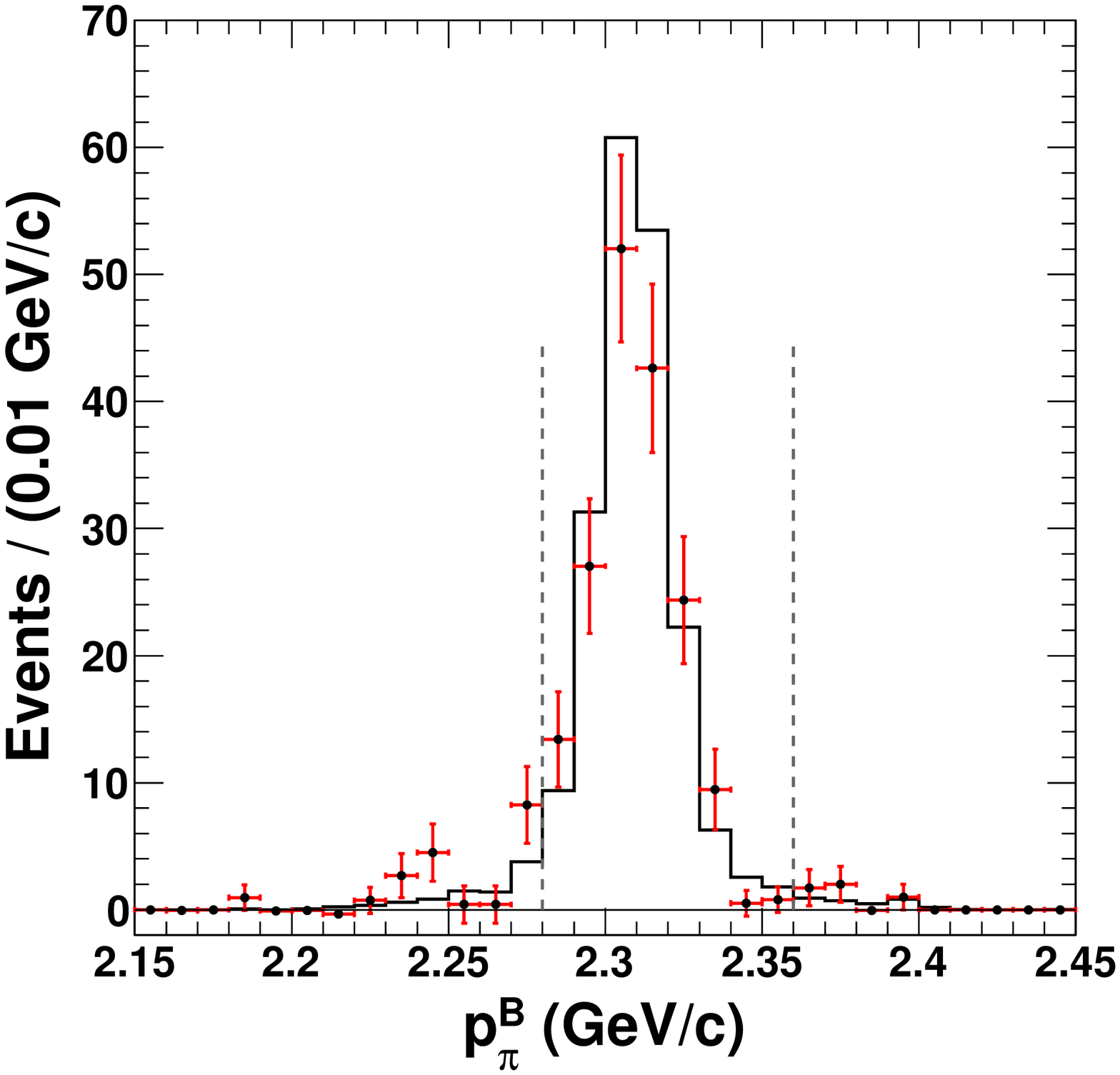}
\caption{The $p_\pi^B$ distributions of the $B^{+}\to \bar{D}^0 \pi^+$
  control sample study. The points with error bars indicate the
  background-subtracted data and the solid histogram shows the MC distribution. 
  The region between the two dashed lines represents the $p_\pi^B$
  selection region for the control sample study.}
\label{control sample}
\end{center}
\end{figure}

To account for the difference of $\mathbf{\it p}_\ell^B$ shapes in the
signal MC and the data, we study $B^+\to \bar{D}^0
\pi^+$ decays as a control sample. The control sample is similar to
our signal decay since it is also a two-body decay of a
$B^+$~meson. The $\bar{D}^0$~meson is identified in the
$\bar{D}^{0}\to K^+ \pi^-$ and $\bar{D}^{0}\to K^+ \pi^- \pi^+ \pi^-$
decay channels. We follow the same analysis procedure as in the $B^+
\to \ell^+ \nu_\ell$ analysis, where the $\pi^+$ from the primary
decay of the $B^+$~meson (primary $\pi^+$), is treated as the lepton
and the $\bar{D}^0$ decay products as a whole are treated as the invisible
neutrino. We compare the distributions of the primary $\pi^+$ momentum
in the rest frame of the signal~$B$ ($p_\pi^B$) between the background
subtracted data sample and the control sample MC, which are displayed
in Fig.~\ref{control sample}.

We estimate the $\mathbf{\it p}_\ell^B$ shape correction factor as the
ratio of the $p_\pi^B$ selection efficiencies between the
background-subtracted data and MC for the control mode.  The yields are
compared for the wide ($2.15~{\rm GeV/}c < p_\pi^B < 2.45~{\rm GeV/}c$)
and the peak ($2.28~{\rm GeV/}c < p_\pi^B < 2.36~{\rm GeV/}c$) region,
separately for data and MC.  By comparing the ratios of the peak region
yield to that of the wide region, we obtain the 
$\mathbf{\it p}_\ell^B$ shape correction factor as $0.953\pm 0.034$, where the
error includes both the statistical uncertainty of the study as well
as systematic uncertainties in fitting. With this correction
applied to the MC sample, the control sample yield of data and MC 
agree within $0.3\sigma$.

The total systematic uncertainty related to $\epsilon_{\rm s}N_{B^+
B^-}$ is $8.0\%$ for both $B^+ \to e^+ \nu_e$ and
$B^+ \to \mu^+ \nu_\mu$. The
multiplicative uncertainties related to $\epsilon_{\rm s}N_{B^+ B^-}$
are summarized in Table~\ref{sys}.

\begin{figure}[htb]
\includegraphics[width=0.45\textwidth]{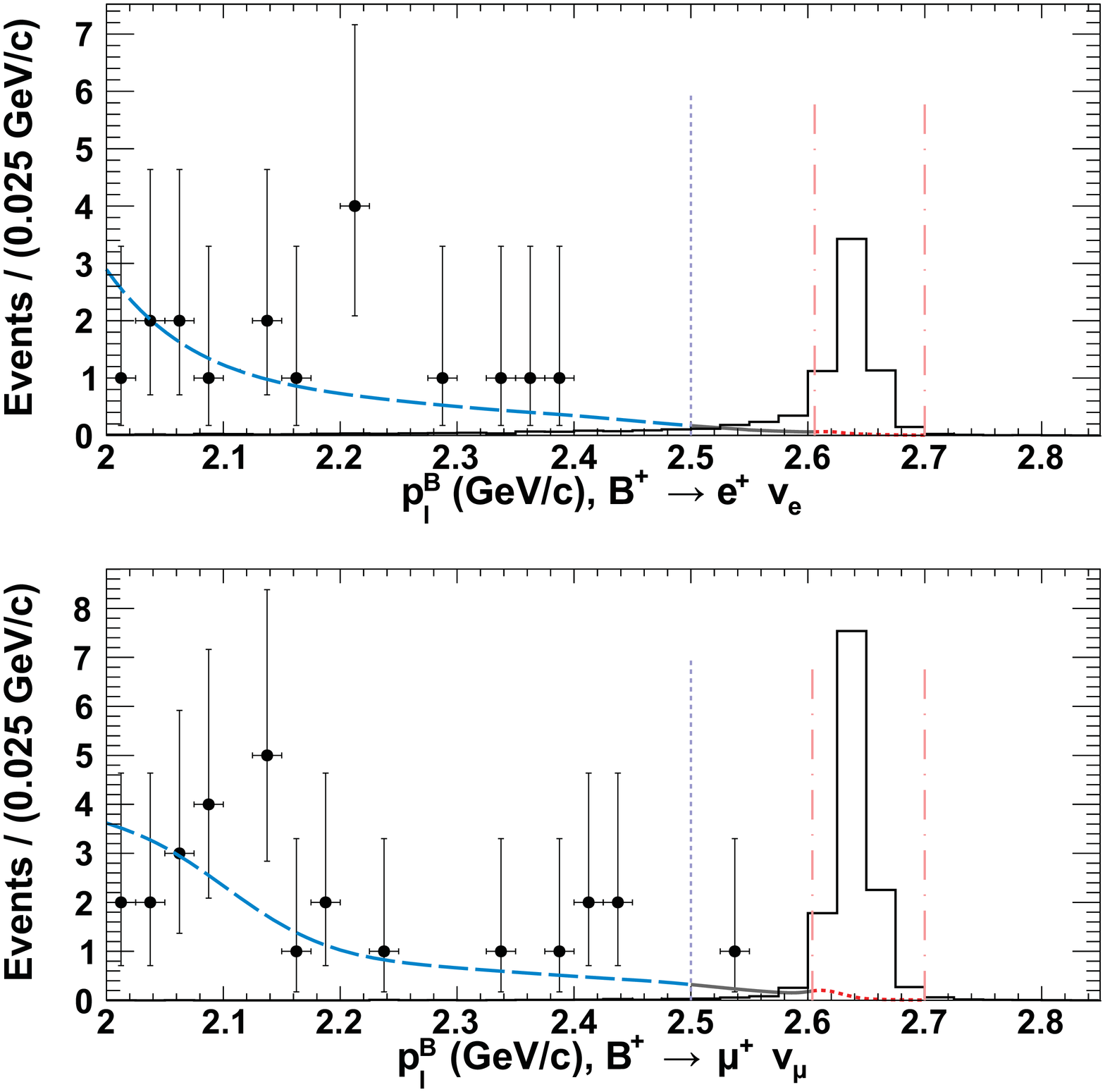}
\caption{The $p_{\ell}^{B}$ distributions for $B^+ \to
    e^+ \nu_e$ (top) and $B^+ \to \mu^+ \nu_\mu$ (bottom). The points
    with error bars are the experimental data.  The solid histograms are
    for the signal MC distributions which are scaled up by a factor of
    $10^6~(40)$ from the SM expectation for $B^+ \to e^+ \nu_e$ ($B^+
    \to \mu^+ \nu_\mu$). The dashed (blue) curves show the background
    PDF fitted in the sideband region ($2.0~{\rm GeV}/c < \mathbf{\it
      p}_\ell^B < 2.5~{\rm GeV}/c$). The vertical dotted line shows
    the upper bound of the $p_\ell^B$ sideband, while the region
    between the two dot-dashed (red) vertical lines correspond to the
    $p_\ell^B$ signal region.}
\label{plB dist}
\end{figure}

In the $\mathbf{\it p}_\ell^B$ signal region, we observe no events for
either search as shown in Fig.~\ref{plB dist}. We set 90$\%$
C.L. branching fraction upper limits using the POLE
program~\cite{POLE} based on a frequentist approach~\cite{FC}. In the
calculation, we assume a Gaussian distribution of $N_{\rm exp}^{\rm
  bkg}$, with a conservative assumption by choosing the larger
deviation of the asymmetric uncertainty in $N_{\rm exp}^{\rm bkg}$. We
obtain upper limits of the branching fraction for each mode as
${\cal B}(B^+ \to e^+ \nu_e)$ $< 3.5\times 10^{-6}$ and ${\cal
  B}(B^+ \to \mu^+ \nu_\mu)<2.7\times 10^{-6}$ at 90$\%$ C.L, which
include the systematic uncertainties.

In summary, we have searched for the leptonic decays $B^+ \to e^+
\nu_e$ and $B^+ \to \mu^+ \nu_\mu$ with the hadronic tagging method
using a data sample containing $772\times 10^6~B\bar{B}$ events
collected by the Belle experiment. We find no evidence of $B^+ \to e^+
\nu_e$ and $B^+ \to \mu^+ \nu_\mu$ processes. We set the upper limits
of the branching fraction at ${\cal B}(B^+ \to e^+ \nu_e)$
$< 3.5\times 10^{-6}$ and ${\cal B}(B^+ \to \mu^+
\nu_\mu)<2.7\times 10^{-6}$ at 90$\%$ C.L, which are by far the most
stringent limits obtained with the hadronic tagging method. Given the
low background level demonstrated in this search, we expect more
stringent constraints on the new physics models to be set by
Belle~II~\cite{B2}, the next generation $B$~factory experiment.


We thank the KEKB group for the excellent operation of the
accelerator; the KEK cryogenics group for the efficient
operation of the solenoid; and the KEK computer group,
the National Institute of Informatics, and the 
PNNL/EMSL computing group for valuable computing
and SINET4 network support.  We acknowledge support from
the Ministry of Education, Culture, Sports, Science, and
Technology (MEXT) of Japan, the Japan Society for the 
Promotion of Science (JSPS), and the Tau-Lepton Physics 
Research Center of Nagoya University; 
the Australian Research Council and the Australian 
Department of Industry, Innovation, Science and Research;
Austrian Science Fund under Grant No. P 22742-N16;
the National Natural Science Foundation of China under Contracts 
No.~10575109, No.~10775142, No.~10825524, No.~10875115, No.~10935008 
and No.~11175187; 
the Ministry of Education, Youth and Sports of the Czech
Republic under Contract No.~LG14034;
the Carl Zeiss Foundation, the Deutsche Forschungsgemeinschaft
and the VolkswagenStiftung;
the Department of Science and Technology of India; 
the Istituto Nazionale di Fisica Nucleare of Italy; 
the WCU program of the Ministry of Education Science and
Technology, National Research Foundation of Korea Grants
No.~2011-0029457, No.~2012-0008143, No.~2012R1A1A2008330,
No.~2013R1A1A3007772;
the BRL program under NRF Grant No.~KRF-2011-0020333,
No.~KRF-2011-0021196,
Center for Korean J-PARC Users, No.~NRF-2013K1A3A7A06056592; the BK21
Plus program and the GSDC of the Korea Institute of Science and
Technology Information;
the Polish Ministry of Science and Higher Education and 
the National Science Center;
the Ministry of Education and Science of the Russian
Federation and the Russian Federal Agency for Atomic Energy;
the Slovenian Research Agency;
the Basque Foundation for Science (IKERBASQUE) and the UPV/EHU under 
program UFI 11/55;
the Swiss National Science Foundation; the National Science Council
and the Ministry of Education of Taiwan; and the U.S.\
Department of Energy and the National Science Foundation.
This work is supported by a Grant-in-Aid from MEXT for 
Science Research in a Priority Area (``New Development of 
Flavor Physics'') and from JSPS for Creative Scientific 
Research (``Evolution of Tau-lepton Physics'').



\begin{thebibliography}{99}

\bibitem{SM}
D.~Silverman and H.~Yao, Phys. Rev. D {\bf 38}, 214 (1988).
\bibitem{CKM}
N.~Cabibbo, Phys. Rev. Lett {\bf 10}, 531 (1963); M.~Kobayashi and T.~Maskawa, Prog. Theor. Phys. {\bf 49}, 652 (1973).
\bibitem{PDG}
J.~Beringer {\it et al.} (Particle Data Group), Phys. Rev. D {\bf 86}, 010001 (2012) and 2013 partial update for the 2014 edition.
\bibitem{FBQCD}
R.~J.~Dowdall {\it et al.} (HPQCD Collaboration), Phys. Rev. Lett. {\bf 110}, 222003 (2013).
\bibitem{taube}
K.~Hara {\it et al.} (Belle Collaboration), Phys. Rev. D {\bf 82}, 071101(R) (2010); K.~Hara {\it et al.} (Belle Collaboration), Phys. Rev. Lett. {\bf 110}, 131801 (2013).
\bibitem{tauba}
B.~Aubert {\it et al.} ($\textsc{Babar}$ Collaboration) Phys. Rev. D {\bf 76}, 052002 (2007); B.~Aubert {\it et al.} ($\textsc{Babar}$ Collaboration), Phys. Rev. D {\bf 77}, 011107(R) (2008).
\bibitem{Belle2007}
N.~Satoyama {\it et al.} (Belle Collaboration), Phys. Lett. B {\bf 647}, 67 (2007).
\bibitem{Babar2009}
B.~Aubert {\it et al.} ($\textsc{Babar}$ Collaboration), Phys. Rev. D {\bf 79}, 091101 (2009).
\bibitem{2HDM}
W.-S.~Hou, Phys. Rev. D {\bf 48}, 2342 (1993).
\bibitem{MSSM}
S.~Baek and Y.~G.~Kim, Phys. Rev. D {\bf 60}, 077701 (1999).
\bibitem{LQ}
H.~Georgi and S.~L.~Glashow, Phys. Rev. Lett. {\bf 32}, 438 (1974).
\bibitem{MLFV}
V.~Cirigliano, B.~Grinstein, G.~Isidori, and M.~B.~Wise, Nucl. Phys. {\bf B728}, 121 (2005).
\bibitem{FI}
A.~Filipuzzi and G.~Isidori, Eur. Phys. J. C {\bf 64}, 55 (2009).
\bibitem{tanbeta}
{The parameter $\tan\beta$ is the ratio of the vacuum expectation values of the two Higgs fields; see A. Djouadi and J. Quevillon, J. High Energy Phys.~10 (2013) 028.}
\bibitem{Lepuniv}
G.~Isidori and P.~Paradisi, Phys. Lett. B {\bf 639}, 499 (2006).
\bibitem{Babar2008}
B.~Aubert {\it et al.} ($\textsc{Babar}$ Collaboration), Phys. Rev. D {\bf 77}, 091104 (2008).
\bibitem{Nuh}
T.~Asaka, S.~Blanchet, and M.~Shaposhnikov, Phys. Lett. B {\bf 631}, 151 (2005); T.~Asaka and M.~Shaposhnikov, Phys. Lett. B {\bf 620}, 17 (2005).
\bibitem{Belle}
A.~Abashian {\it et al.} (Belle Collaboration), Nucl. Instrum. Methods Phys. Res. Sect. A {\bf 479}, 117 (2002); also see detector section in J.Brodzicka {\it et al}., Prog. Theor. Exp. Phys. (2012) 04D001.
\bibitem{KEKB}
S.~Kurokawa and E.~Kikutani, Nucl. Instrum. Methods Phys. Res. Sect. A {\bf 499}, 1 (2003), and other papers included in this volume; T.Abe {\it et al.}, Prog. Theor. Exp. Phys. (2013) 03A001 and following articles up to 03A011.
\bibitem{LID}
K.~Hanagaki {\it et al.}, Nucl. Instrum. Methods Phys. Res., Sect. A {\bf 485}, 490 (2002); A.~Abashian {\it et al.}, Nucl. Instrum. Methods Phys. Res., Sect. A {\bf 491}, 69 (2002).
\bibitem{EKP}
M.~Feindt {\it et al.}, Nucl. Instrum. Methods Phys. Res., Sect. A {\bf 654}, 432 (2011).
\bibitem{Ulnu}
A.~Sibidanov {\it et al.} (Belle Collaboration), Phys. Rev. D {\bf 88}, 032005 (2013).
\bibitem{Barlow}
R.~Barlow and C.~Beeston, Comput. Phys. Comm. {\bf 77} (1993) 219-228.
\bibitem{Lnugam}
G.~Korchemsky, D.~Pirjol and T.-M.~Yan, Phys. Rev. D {\bf 61}, 114510 (2000). 
\bibitem{POLE}
J.~Conrad {\it et al.}, Phys. Rev. D {\bf 67}, 012002 (2003).
\bibitem{FC}
G.~J.~Feldman and R.~D.~Cousins, Phys. Rev. D {\bf 57}, 3873 (1998).
\bibitem{B2}
T.~Abe {\it et al.} (Belle Collaboration), 	arXiv:1011.0352v1 [physics.ins-det] (2010).


\end{thebibliography}
\end{document}